\documentclass[useAMS,usenatbib]{mn2e}

\newcommand\mone{$^{-1}$}
\newcommand\mtwo{$^{-2}$}

\newcommand{\oo}{\rm [O\,II]$\lambda\lambda$3727,3729/[O\,III]$\lambda$5007}

\usepackage{graphicx}
\usepackage{txfonts}
\title[First X-ray observations of Low-Power Compact Steep Spectrum Sources]
  {First X-ray observations of Low-Power Compact Steep Spectrum Sources}

\author[M.~ Kunert-Bajraszewska et al.]
{M.~ Kunert-Bajraszewska$^1$, A.~Labiano,$^{2,3}$, A.
Siemiginowska$^3$, M.~Guainazzi,$^2$ \\
$^1$ Toru\'n Centre for Astronomy, Faculty of Physics, Astronomy
and Informatics, NCU, Grudziacka 5, 87-100 Toru\'n, Poland\\
$^2$European Space Astronomy Centre of ESA, PO Box 78, Villanueva de la Ca\~nada, 28691, Madrid, Spain\\
$^3$Harvard-Smithsonian Center for Astrophysics, Cambridge, MA 02138, USA}
\date{}

\pagerange{\pageref{firstpage}--\pageref{lastpage}} \pubyear{2013}

\def\LaTeX{L\kern-.36em\raise.3ex\hbox{a}\kern-.15em
    T\kern-.1667em\lower.7ex\hbox{E}\kern-.125emX}

\begin{document}
\label{firstpage}

\maketitle
\begin{abstract}
We report first X-ray {\it Chandra} observations of a sample of seven low
luminosity compact (LLC) sources. 
They belong to a class of young compact steep spectrum (CSS) radio sources.
Four of them have been detected, the other three have upper limit estimations 
for X-ray flux, one CSS galaxy is associated with an X-ray cluster. 
We have used the new observations together with the observational data for known strong CSS
and gigahertz-peaked spectrum (GPS) objects and large scale FR\,Is and FR\,IIs to study the relation
between morphology, X-ray properties and excitation modes in radio-loud AGNs.  
We found that: (1) The low power objects fit well to the already established
X-ray - radio luminosity correlation for AGNs and occupy the space among, weaker in
the X-rays, FR\,I objects. (2) The high excitation galaxies (HEG) and low
excitation galaxies (LEG) occupy
distinct locus in the radio/X-ray luminosity plane, notwithstanding their  
evolutionary stage. 
This is in agreement with the postulated different origin of the X-ray
emission in these two group of objects.
(3) We have tested the AGN evolution models
by comparing the radio/X-ray luminosity ratio with the size of the sources, and indirectly, with
their age. We conclude that the division for two different X-ray emission
modes, namely originate in the base of the relativistic jet (FR\,Is) or in
the accretion disk (FR\,IIs) is already present among the younger compact
AGNs. (4) Finally, we found that the CSS sources are less obscured than the
more compact GPSs in X-rays. However, the anti-correlation between X-ray column
density and radio size does not hold for the whole
sample of GPS and CSS objects.

\end{abstract}



\section{Introduction}

We still know little about how radio galaxies are born and how they subsequently evolve, 
but it is generally accepted that the GHz Peaked Spectrum (GPS) and Compact Steep Spectrum 
(CSS) radio sources are young, smaller versions of the large-scale powerful radio sources 
\citep{O'Dea91,Fanti90,Fanti95,Readhead96a,O'Dea97}. Recently, the High
Frequency Peakers have been added to the sequence, as possible
progenitors of GPS sources \citep[e.g., ][ and references
therein]{Orienti07}.

The GPS and CSS sources are powerful but compact radio sources whose spectra are generally 
simple and convex with peaks near 1 GHz and 100 MHz respectively. The GPS sources are contained 
within the  extent of the optical narrow emission line region ($\lesssim 1$ kpc) while the 
CSS sources are contained within the host galaxy \citep[$\lesssim 15$ kpc, see ][ for a review]{O'Dea98}.

In the general scenario of the evolution of powerful radio-loud AGNs, 
GPS sources evolve into CSS sources and these into supergalactic-size 
FR\,I or FR\,II objects \citep{Fanaroff74}. The dynamic evolution of the
double-lobed radio sources, characterized by the total extent of the source,
advance speed of the hotspots and the dependence of the density distribution
of the interstellar and intergalactic medium along the way of the
propagating jets and lobes, predicts the increase of the radio power with
the linear size of the source in the GPS and CSS phase until they {\bf
reach} the
1-3\,kpc size. Then the larger CSS objects should start to slowly decrease
their luminosity but the sharp radio power decrease is visible only in the FR\,I
and FR\,II phase of evolution \citep{Begelman89, Young93, Carvalho94,
Fanti95, Begelman96, Young97, Kaiser97b, Carvalho98, Snellen00, Kino05, Kawakatu06, Kaiser07,
Kawakatu09a}. Finally, after the cut off of the material supply to the
central engine of the galaxy, the sources begin their fading phase. They can
come back on the main evolutionary sequence after the re-ignition of the
radio activity \citep[e.g., ][]{koziel12, konar12}.


\begin{table*}
\caption[]{Basic properties of the sample and X-ray models. Redshifts
followed by a
``p'' are photometric. Fluxes are in $10^{-15}$ erg s\mone cm\mtwo. Limits
are 3$\sigma$. The numbers in
parentheses indicate the errors calculated as $\sqrt{counts}$.}
\label{observations}
\resizebox{\textwidth}{!}{
\begin{tabular}{cccccccccccc}
\hline
Source   & RA(J2000) & Dec (J2000) & ID & {\it z}  & 
Counts& F$_{0.5-2 \rm{keV}}$ & F$_{2-10 \rm{keV}}$ &
N$_{H, Gal}$ & N$_H$ & $\Gamma$& {\it Chandra}\\
&&&&&${0.5-7 \rm{keV}}$& & & ($10^{20}\, cm^{-2}$)& ($10^{20}\, cm^{-2}$) &
&Obs ID\\
\hline 

{0810+077}  & 08:13:23.76 & 07:34:05.80& Q & 0.112 & 119 (11) &
32$^{+11}_{-10}$ & 190$^{+140}_{-90}$ &
2.14& 13$^{+1.3}_{-1.3}$  &  0.64$^{+0.15}_{-0.15}$& 12716\\

{0907+049}  &09:09:51.13 & 04:44:22.13& G & 0.640p & 3$^{a}$   &
$<$2.5 & $<$4.6& 1.41 & $-$& 1.7$^{b}$& 12717\\ 

{0942+355}  &09:45:25.89 & 35:21:03.50& G & 0.208 & 103 (10) &
32$^{+5.0}_{-5.0}$ & 66$^{+33}_{-24}$ &
3.55& $<$23  &  1.6$^{+0.24}_{-0.16}$& 12714\\ 

{1321+045} &13:24:19.70 & 04:19:07.20& G & 0.263 & 53 (7) &
14$^{+3.0}_{-3.0}$ & 9.4$^{+16.6}_{-5.4}$ &
2.04& $-$  & 2.35$^{+0.39}_{-0.36}$& 12715\\  

{1542+390} &15:43:49.49 & 38:56:01.40& G & 0.553 & 0$^{a}$ &
$<$2.5 & $<$4.6 & 1.55&  $-$& 1.7$^{b}$& 12718\\  

{1558+536}  &15:59:27.66 & 53:30:54.70& G & 0.179 & 9 (3) &
2.8$^{+1.5}_{-1.5}$ & 5.2$^{+1.6}_{-1.6}$  &
1.26&  $-$&  1.7$^{b}$& 12719\\ 

{1624+049}  &16:26:50.30 & 04:48:50.50& G & 0.040p  & 4$^{a}$ &
$<$1.2 &  $<$2.6 & 5.01  &
 $-$& 1.7$^{b}$& 12720 \\

\hline
\end{tabular}
}
\begin{minipage}{160mm}
$^{a}$ means no detection, only upper limit for flux.\\
$^{b}$ means $\Gamma$=1.7 was assumed for the flux calculation. 
\end{minipage}
\end{table*}

\begin{figure*}
\centering
\includegraphics[width=\columnwidth]{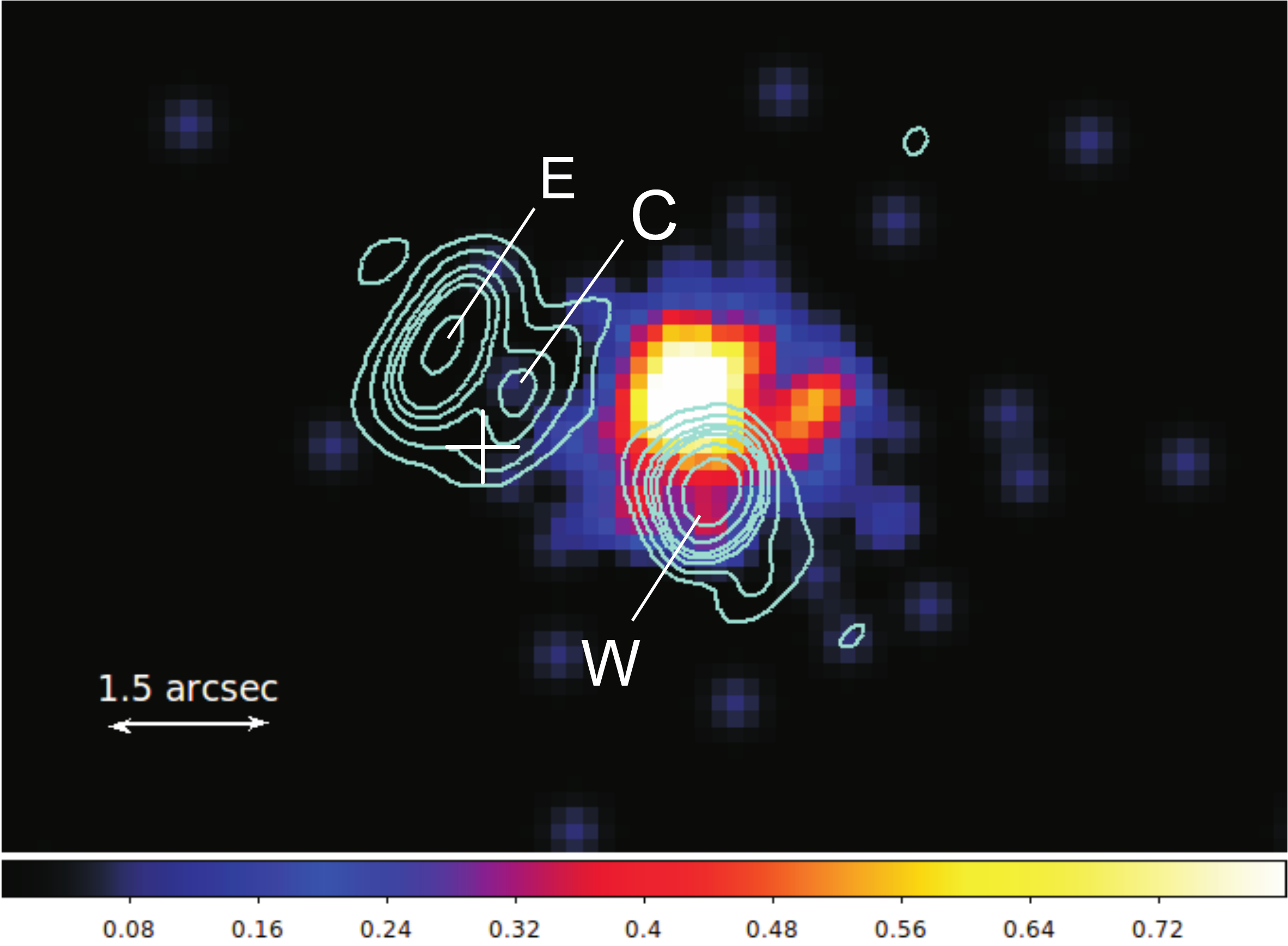} \hfill
\includegraphics[width=\columnwidth]{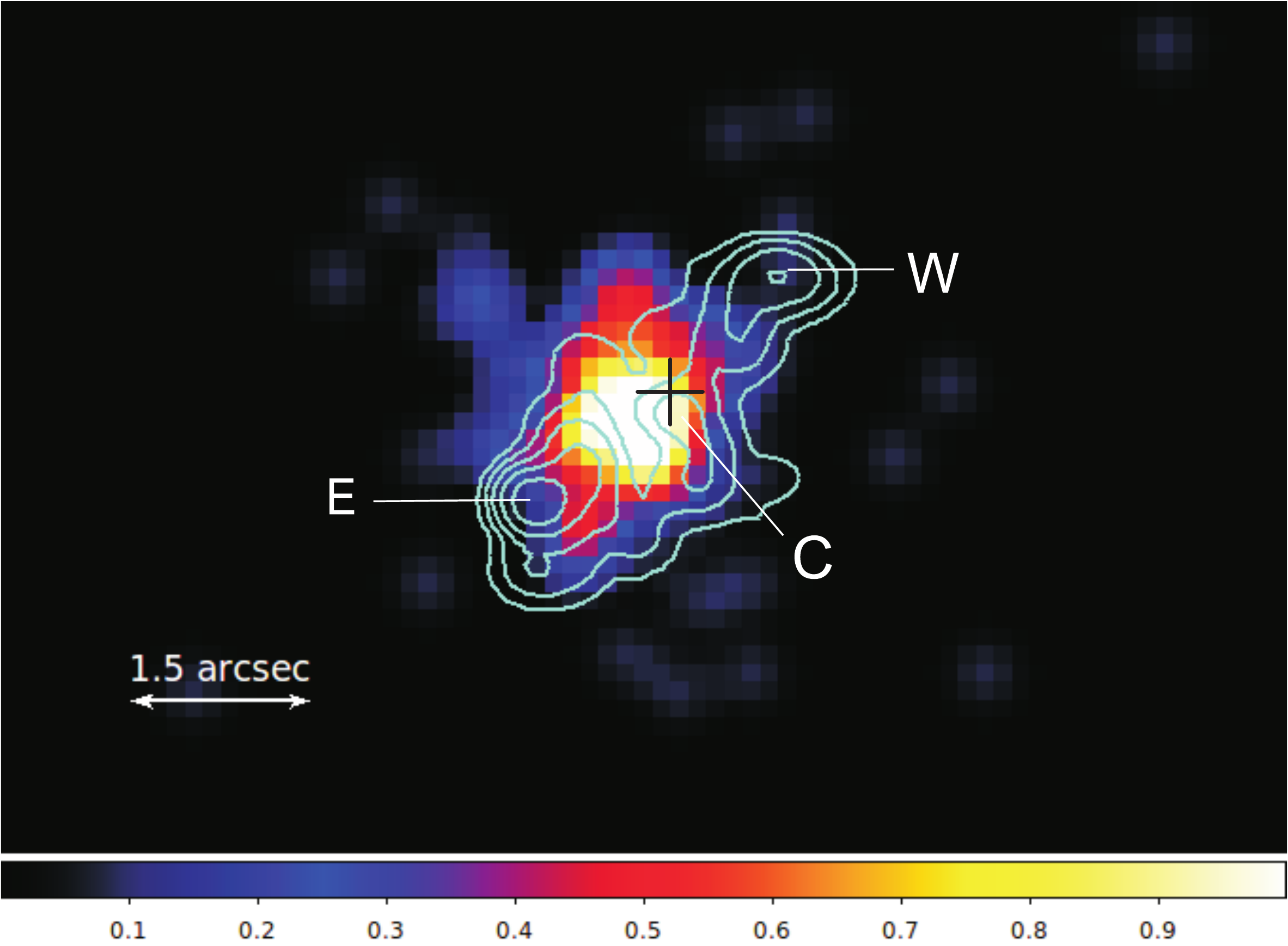}
\caption{{\it Chandra} X-ray (color) and MERLIN 1.6\,GHz (contours) emission
from 0810+077 (left panel) and 0942+355 (right panel). The radio maps are taken 
from \citep{Kunert10a}. We have used 0.15 pixel blocking for the X-ray images.
The cross indicates the position of an optical counterpart taken from the
SDSS. The radio contours increase by factor of 2, the first contour level   
corresponds to $\approx3\sigma$ and amounts 1.7 mJy/beam (0810+077) and 0.85
mJy/beam (0942+355).} 
\label{oplots}
\end{figure*}

However, population studies have drawn attention to the existence of far too many
compact sources compared to the number of large-scale objects
\citep{O'Dea97}. It has been proposed then that some of the young radio-loud
AGNs, namely the GPS and CSS
sources, can be short-lived objects \citep{rb97, czerny, Kawakatu09b,Kunert10a} and that not one but a few
evolutionary paths exist \citep{Marecki03, Kunert10a, an12}. Detection of several candidates for dying
compact sources \citep{Giroletti05, Kunert06, Kunert10a, orienti} supports this view.
The determining factors for the further evolution of compact radio objects could occur at 
sub-galactic (or even nuclear) scales, or they could be related 
to the radio jet-ISM interactions and evolution. Our previous studies suggest that the evolutionary 
track could be related to the interaction, strength of the radio source, and excitation levels of 
the ionized gas \citep{Kunert10a, Kunert10b}, instead of the radio morphology of the young 
radio source. 

The characteristics (size, radio power and young age) of GPS and CSS sources make them excellent 
probes of interaction (and therefore evolution) of radio sources. Furthermore, they have not 
completely broken through the ISM, so these interactions are expected to be more important than 
in the larger sources. Observations of UV, HI and, especially, of the ionized gas in GPS and CSS 
sources suggest the presence of such interactions 
\citep[][]{Labiano08letter, Labiano08, Holt06, Labiano05, Axon00, Vries99, Vries97, Gelderman94}.

Additional clues on the evolution of compact GPS and CSS sources may come
from the X-ray band, but still little is known about the nature of the X-ray emission in
these young sources. Theoretical models predict 
strong X-ray emission from young radio sources, due to the recent triggering
of the nuclear activity
and/or the expansion through the ISM \citep[e.g., ][and references therein]
{Siemiginowska2012,Siemiginowska09}.
The {\it Chandra} and {\it XMM-Newton} observations of GPS and CSS objects
made so far have focused on sources with high
radio emission \citep[e.g., ][]{Siemiginowska08, Vink06, Guainazzi04,
Guainazzi06, Tengstrand09}. These sources, when included in the L$_{2-10}$
keV versus L$_{5GHz}$ diagram, group in the region occupied by powerful
FR\,II sources \citep{Tengstrand09}.
Therefore, the location of GPS and CSS sources in the
radio to X-ray luminosity
diagram is consistent with them being powered by accretion, and
therefore evolving onto a track of constant X-ray, accretion-driven 
luminosity to FR\,IIs, as well as with the correlation between radio and
X-ray luminosity observed in FR\,Is, which would point to a common
origin for the emission in these two bands.
   
In this paper, we present the first X-ray observations of low power radio sources, starting to fill 
the gap in the  L$_{2-10}$ keV versus L$_{5GHz}$ diagram, and shedding some light on the origin of 
high-energy emission of young radio sources and their evolution.

\section{The sample}

The current sample consists  of 7 sources ({0810+077}, {0907+049}, {0942+355}, {1321+045}, {1542-390}, 
{1558+536}, {1624+049}) taken from the Low Luminosity Compact sources sample 
\citep[LLC, ][]{Kunert09, Kunert10a}. The LLC consists of 44 nearby (z$<$0.9) sources, selected from 
the final release of FIRST \citep{white1997}, and the GB6
\citep{gregory1996} and SDSS surveys, and observed with MERLIN at L-band and 
C-band. The main selection criterion for the LLC was the luminosity limit: 
L$_{5GHz} < 5\times 10^{42}$ erg s\mone. The radio and optical properties of the LLC were  
discussed and analyzed in \citet{Kunert10a} and \citet{Kunert10b} respectively.

The 7 current sources form the so called {\it pilot sample} and  were selected to represent different 
stages of the radio source evolution 
within the ISM: weak or undetected radio core and strong lobes or breaking up radio lobes with 
bright radio core, and linear sizes ranging from 2 to 17 kpc .

\section{X-ray observations and data reduction}

The sample was observed using  {\it Chandra} ACIS-S3 with 1/8 subarray and standard pointings, 
with exposure times of $\sim$ 9500 seconds (see Table \ref{observations}). The {\it Chandra} 
data were reduced using CIAO 4.5 \citep{Fruscione06} with the calibration files from 
CALDB  4.4.5. All our sources are contained within the FWHM of the PSF. We used a circular 
extraction region for each source, with radius $2\arcsec$, which also contains all the 
radio emission. The background regions consist of four circular regions of radius 10$\arcsec$ 
around the source. The CIAO default tools were used to extract the spectra and associated 
rmf and arf files. The total counts detected for each source are listed in Table  \ref{observations}.

We used Sherpa \citep{Freeman01} to fit the spectra, using an absorbed power-law in the 0.5-7 keV energy range:

\begin{equation}
\hfill   N(E) =  e^{-N_H^{Gal} \sigma(E)  } \times e^{-N_H^{z_{obs}} \sigma[E(1+ z_{obs})] }  \times AE^{-\Gamma}  \hfill
\end{equation}

where N(E) is in photons  cm$^{-2}$ s$^{-1}$,  A is the normalization at 1 keV, $\Gamma$ is the 
photon index of the power law, $\sigma$(E) and $\sigma$[E(1+ z$_{obs}$)] are the absorption 
cross-sections \citep{Morrison83, Wilms00}, and N$_H^{Gal}$ and N$_H^{z_{obs}}$ are the column 
densities of the Milky Way \citep{Kalberla05, Dickey90} and the source. The Galactic absorption 
was kept constant during fitting.  The second absorption component is assumed to be intrinsic to the 
quasar and located at the redshift of the source. The model was applied to all sources. However, 
{0907+049}, {1558+536} and {1624+049} do not have enough counts to produce a reasonable fit. 
The results are summarized in Table \ref{observations}.

We use H$_0$=71, $\Omega_M = 0.27 ,  \Omega_\Lambda = 0.73$ \citep{Spergel03} throughout the paper.


\begin{table}
\caption{Spectral type and luminosities of the sources.}
\label{lumins}
\begin{tabular}{ccccccccccc}
\hline
Source   & log L$_{2-10 \rm{keV}}$ & log L$_{5 \rm{GHz}}$ & log L$_{\rm[{O}{III}]}$ & Spectral\\
& & & &type\\
\hline 
{0810+077} & 42.8    & 41.4 & 40.8     & LEG \\
{0907+049} & $<$42.9 & 42.6 & $<$42.2  & -- \\ 
{0942+355} & 42.9    & 41.4 & 42.0     & HEG \\ 
{1321+045} & 42.3    & 41.6 & 40.3     & LEG \\  
{1542+390} & $<$42.7 & 42.4 & 41.9     & HEG \\  
{1558+536} & $<$41.7 & 41.4 & 40.8     & LEG \\ 
{1624+049} & $<$40.0 & 40.0 & --       & -- \\
\hline
\end{tabular}
\begin{minipage}{\columnwidth}
Luminosities in erg s\mone. Limits are 3$\sigma$. Spectral type and radio morphology according to 
\citet{Kunert10a, Kunert10b}. 
\end{minipage}
\end{table}


\begin{table}
\caption{Number of ionizing photons.}
\label{photons}   
\centering
\begin{tabular}{ccccc}
\hline
Source   & Distance    & Log N$_{H\beta}$  & Log N$_{Nuc}$ &
N$_{Nuc}$/N$_{H\beta} $\\
\hline
{0810+077}   & 513.6   & 52.7 &  54.7 & 100.0 \\
{0942+355}   & 1014.3  & 53.5 &  55.3 & 63.0 \\ 
{1321+045}   & 1323.9  & 53.0 &  55.1 & 126.0 \\
\hline
\end{tabular}
\begin{minipage}{\columnwidth}
Source name, luminosity distance in Mpc, N$_{H\beta}$ - number of photons/s needed to ionize
H$_\beta$, N$_{Nuc}$ - number of ionizing photons/s produced by the nucleus, and the ratio between the last
two.
If the ratio is $\geq$1, the nucleus is producing enough photons to ionize
H$_\beta$,
if the ratio is lower than one, another source of ionization, such as
shocks, is required.
\end{minipage} 
\end{table}

\section{Discussion}

\subsection{The X-ray and radio morphology}

We have observed a {\it pilot sample} of Low Luminosity Compact (LLC)
sources (7 out of 44 objects) with {\it Chandra}. 
Four of them have been
detected, the other three have upper limit estimations for X-ray flux (see Table
\ref{observations}). One of the objects, 1321+045, appeared to be associated
with an X-ray cluster and has been discussed in a separate paper
\citep{kun2013}.
The {\it Chandra} ACIS-S images of two of the sources discussed here with the largest 
number of X-ray photons are shown in Fig. \ref{oplots}. We also overlayed
the radio MERLIN 1.6\,GHz contours on the X-ray emission with the
indications of radio components. 

0810+077 is a quasar classified as LEG. 
Its radio morphology consist of three components:
the weak central one (C) and two jets/lobes (E and W). The 
optical counterpart is coincident with the component C and we suggested this
could be a radio core \citep{Kunert10a}. However, this is based on
observations at only one radio frequency so it should be treated as tentative. The
brightest part of the X-ray emission lies between the components C and W.
The potential offset between the centroid of the X-ray emission and
component C or W can be consistent with the astrometric
uncertainty of {\it Chandra}. 

0942+355 is a galaxy classified as HEG, larger than 0810+077 and with more
complex radio structure. Its 1.6\,GHz
asymmetric radio morphology consist of three components: weak radio core (C)
and two lobes (E and W). There are also 5\,GHz observations of this source
\citep{Kunert10a}
showing only emission from the south-eastern radio lobe. 
In the case of 0942+355 the brightest part
of the X-ray source is right in the center of the source, between the two
jets.   

1558+536 is a galaxy classified as LEG with diffuse, double-like morphology
\citep{Kunert10a}. Only nine counts were detected in {\it Chandra}
observations of this source and we did not produced an image of it.

The radio and optical properties of the whole sample of LLC sources have been 
discussed and analyzed
by us \citep{Kunert10a, Kunert10b}. We suggested that they can
represent a population of short-lived objects and undergo the CSS phase
of activity many times before they become large scale FR\,I or FR\,II
\citep{Kunert10a, Kunert10b}. What is more, the evolution of the radio
source seems to be independent from its radio morphology but rather determined 
by the properties of the
central engine: strength, accretion mode, excitation level of the
ionized gas. In some objects the surrounding environment could be also an
important factor influencing the evolution \citep{ceglowski}.

Optically many of the LLC sources belong to the class
of low excitation galaxies (LEGs), which are thought to be powered by 
the accretion of hot gas \citep{hard07, butti10} and can be
progenitors of large scale LEGs \citep{Kunert10b}.

\subsection{The optical-line emission}

\citet{Labiano08letter} found that compact AGNs show a strong correlation between 
[O\,III]$\lambda$5007 line luminosity and size of the radio source
suggesting
a possible deceleration in the jet as it crosses the host ISM.
However this correlation breaks up when including LLC sources, more
specifically the compact LEGs \citep{Kunert10b}. As we have already shown in
the optical analysis of the whole sample of LLC sources the LLC HEGs show a
$\sim10$ times higher [O\,III]
$\lambda$5007 luminosities than LLC LEGs. This could be caused by a stronger
jet contribution to the ionization of the ISM in HEGs and/or
indicates differences in the environment of HEG and LEG objects. 

In the pilot sample of LLC sources observed with {\it Chandra}
we have found that two sources, 0810+077 and 0942+355, have the same radio and X-ray
luminosities (Table \ref{lumins}). However, 0942+355 (HEG) has higher [OIII] emission
than 0810+077 (LEG). 
If we compare the OII/OIII y OIII/Hb of the detections with
photoionization and shock models 
\citep[MAPPINGS,][]{Allen08}, 0810+077 is consistent
with 100\% photoionization and 0942+355 with 80\% photoionization and
20\% shocks  \citep{Kunert10b}.

\begin{figure*}
\centering
\includegraphics[width=8.5cm, height=6cm]{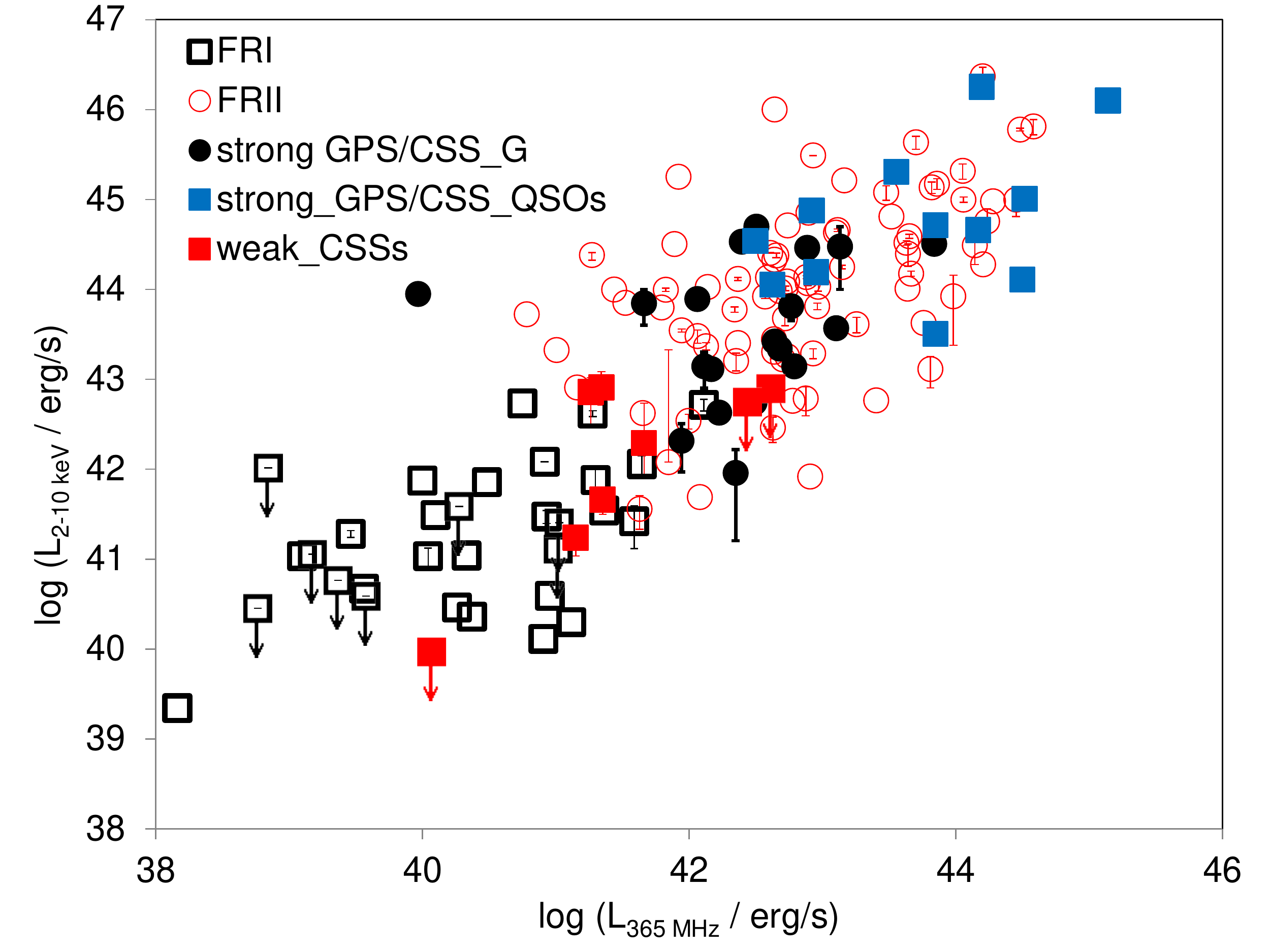}
\hfill
\includegraphics[width=8.5cm, height=6cm]{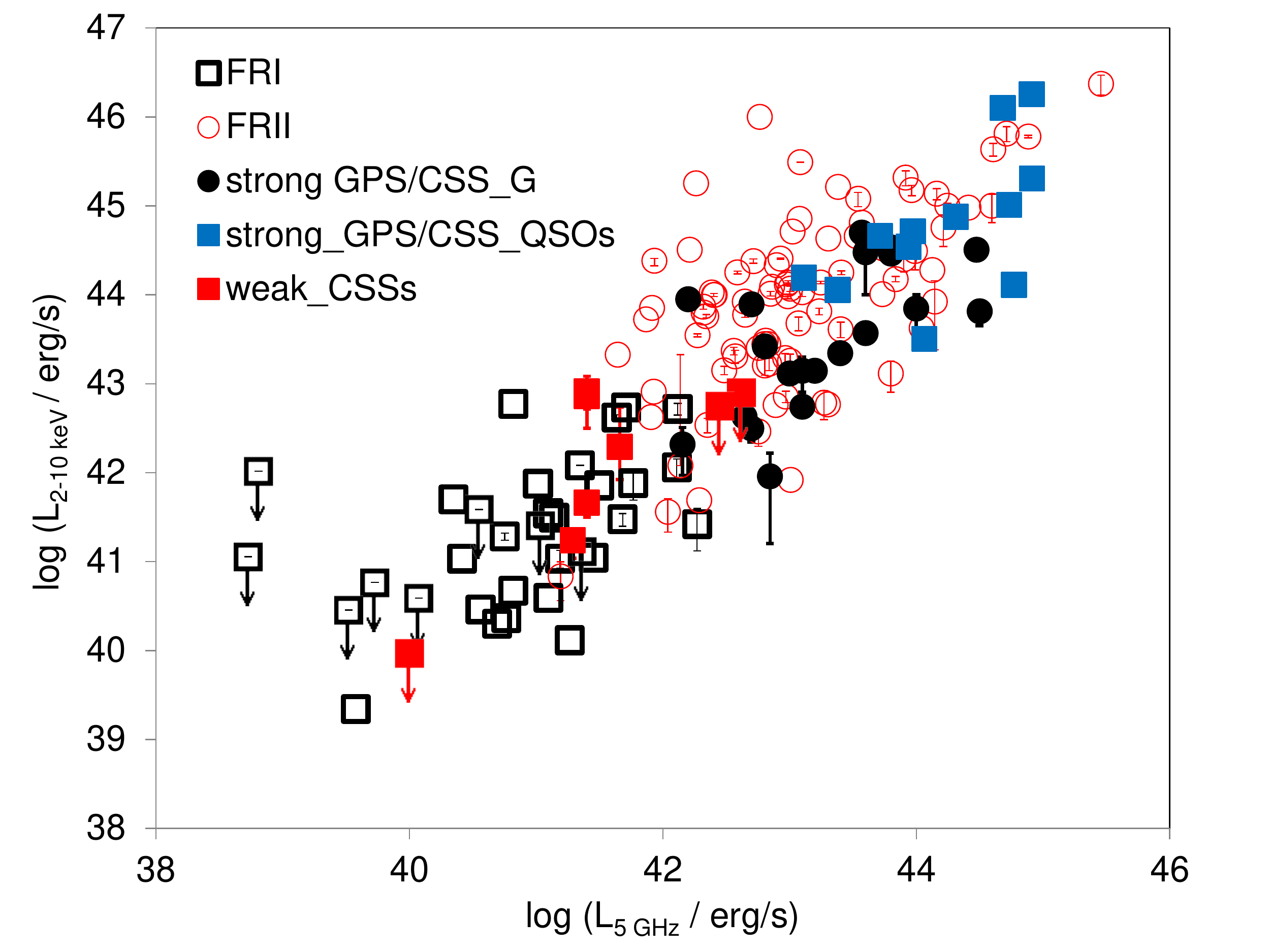}
\caption{Luminosity diagrams for AGNs: 2-10\,keV - 365\,MHz (left) and 
2-10\,keV - 5\,GHz (right). Open red circles
indicate FR\,II sources and black open squares - FR\,Is. Strong GPS and CSS
galaxies and quasars were plotted separately as black circles and blue
squares respectively. Weak CSS sources are indicated with red squares.}
\label{5GHz_plot}
\end{figure*}

\subsection{Can the central AGN power the emission line luminosity in the extended nebulae?}

We compared the number of ionizing photons produced by the nucleus of the source, with the 
number of photons needed to produce the observed emission line luminosity 
\citep[see e.g.][]{Wilson88,Baum89,Axon00,O'Dea00}. Assuming radiative recombination 
under case B conditions, the number of ionizing photons/s, $N_\mathrm{H\beta}$, needed 
to produce the observed H$\beta$ luminosity $L_\mathrm{H\beta}$ is:

\begin{equation}
\hfill N_\mathrm{H\beta} = 2.1\times 10^{52} ( L_\mathrm{H\beta}  / 10^{40} \mathrm{erg \, s}^{-1} ) \hfill 
\end{equation}

We use the integrated [{O}{III}]$\lambda5007$ fluxes from \citet{Kunert10b} 
and scale using the typical ratio for the narrow 
line components in CSS sources \citep[e.g.][]{Gelderman94}: $H\beta/[{O}{III}]~\lambda5007 = 0.18 \pm 0.02$. 

The number of photons/s in the continuum, between frequencies $\nu_1$ and $\nu_2$ is given by:

\begin{equation}
\hfill N_\mathrm{Nuc} = 4\pi D^2 S_\mathrm{0} (\alpha h)^{-1}\ (\nu _1^{-\alpha} - \nu _2^{-\alpha}) \hfill
\end{equation}

where D is the luminosity distance, the flux density spectrum is given by $F_\nu=S_0\nu^{-\alpha}$ 
\citep[we adopt $\alpha$=1, e.g. ][]{O'Dea00} and h is Planck's constant. We are only 
interested in the photons with enough energy to ionize Hydrogen, i.e. those between 
$\nu _1$ = 3.3x10$^{15}$Hz (912\AA ~or 13.6eV) and $\nu _2$ = 4.8x10$^{17}$Hz (2 keV). 
For our spectral index, $\alpha$=1, higher frequencies do not add a significant  number 
of photons. Note that this analysis is subject to the caveat that the continuum emission 
may not be emitted isotropically, and the extended nebulae may see a different luminosity 
than we do \citep[e.g.][]{Penston90} \\

The results are shown in Table \ref{photons}. We find  that the nucleus apparently 
produces enough ionizing photons to power the emission line luminosity in {0810+077} 
and {0942+355}. This is consistent with the results from the comparison with the BPT 
diagrams and MAPPINGS models \citep{Kunert10b, Allen08}.

\subsection{Radio/X-ray correlations}

The {\it Chandra} and {\it XMM-Newton} studies of GPS/CSS sources performed
so far show they are strong X-ray emitters \citep{Guainazzi04, Guainazzi06,
Vink06, Siemiginowska08, Tengstrand09}. 
The X-ray emission of CSS and GPS objects is
probably a
result of a recent triggering of the nuclear activity and can be
characterized by an absorbed power law model with high ($>10^{22}~{\rm
cm^{-2}}$) column densities \citep{Guainazzi06,Vink06,Siemiginowska08,
Tengstrand09}.
But there are also several detections of X-ray morphology in these
compact objects.
Extended hot 0.5-1\,keV interstellar medium (ISM) has been detected in the
case of two CSS sources, 3C303.1 \citep{odea2006} and 3C305
\citep{messi2009}, and it is interpreted as shock-heated environment gas.
The X-ray jets in GPS sources and large scale X-ray emission associated with
some of them have been reported by \citep{Siemiginowska08}. These
objects are classified as 'GPS sources with extended emission' and discussed
in the frame of the theory of intermittent radio activity \citep{stan05}.   
However, the radio structures of GPS and CSS sources are much smaller than
the
spatial resolution of the current X-ray instruments in most cases, what
prevents us from identification of the origin of their X-ray emission.
There are several theoretical predictions of X-ray emission from evolving
radio sources: i) as thermal emission emitted by the ISM of
the host galaxy shock heated by the expanding radio structure
\citep{Heinz98, odea2006}, or ii) that produced in the accretion disk's hot
corona \citep{Guainazzi04, Guainazzi06, 
Vink06, Siemiginowska08}, and finally iii) as non-thermal radiation produced
through IC scattering of the local thermal radiation fields off the lobe
electron population \citep{Stawarz08, Ostorero10} or by mini shells
\citep{Kino13}.   
\citet{Tengstrand09} show that the radio
versus X-ray luminosity plane can be a useful tool to derive constraints on
the evolution of compact
radio sources. Studies of compact radio AGN so far have been biased towards
high-luminosity objects
($L_{5 GHz} > 10^{42}$~erg~s$^{-1}$). In this Section we extend these
studies to the low-luminosity
regime that our pilot {\it Chandra} study probes for the first time.

Our goal is to compare the X-ray properties of different groups of
radio objects, GPS, CSS and large-scale FR\,I and FR\,II sources as well as
the X-ray properties of low and high power compact AGNs. For this purpose we
have built the control sample of GPS/CSS sources \citep{Siemiginowska08,
Tengstrand09, Massaro10, Massaro12} and FR\,I and FR\,II objects
\citep{Sambruna99, Donato04, Grandi06, Evans06, Balmaverde06, Belsole06,
hard06, Massaro10, Massaro12} from results recently published in the
literature. Our pilot sample of low luminosity CSS sources consist of
only seven objects. That
is why we have also included in it the low luminosity 3C305
described by \citet{messi2009}. The total number of GPS/CSS sources is 40
objects, and FR\,I and FR\,IIs - 34 and 85 sources, respectively.
The samples are, however, biased in terms of
their redshift distribution. The GPS/CSS and FR\,II samples are well matched in the redshift,
but the FR\,Is are generally at lower redshift. There are 6 GPS/CSS objects
with redshift in the range $1>{\it z}<2$. All other sources from all groups have
redshit ${\it z}<1$. 

For all plots presented in this paper we have used the total radio and
X-ray luminosity in case of all group of sources. The
reason for that is a lack of information of radio core fluxes of most of our   
compact GPS and CSS sources. Among the seven low power CSS objects only 
two have 5\,GHz observations but without core detection \citep{Kunert10a}.  
Significant part of our sample of GPS/CSS sources is also unresolved in
X-rays. The exceptions from the above-stated rule are a few GPS sources with
extended structures \citep{stan05}. In the case of them the radio and X-rays
values used in this paper refers to their milliarcseconds VLBA structures as
reported in \citet{Tengstrand09} and \citet{Siemiginowska08}.

\subsubsection{Radio/X-ray luminosity plane}

We have compared the X-ray luminosity of the sources from the pilot sample with their radio
properties at 5\,GHz and 365\,MHz (Fig.~\ref{5GHz_plot}).
We have included also the control sample of GPS/CSS and FR\,I and FR\,II
objects as described above.


\begin{figure}  
\centering
\includegraphics[width=8.5cm,
height=6cm]{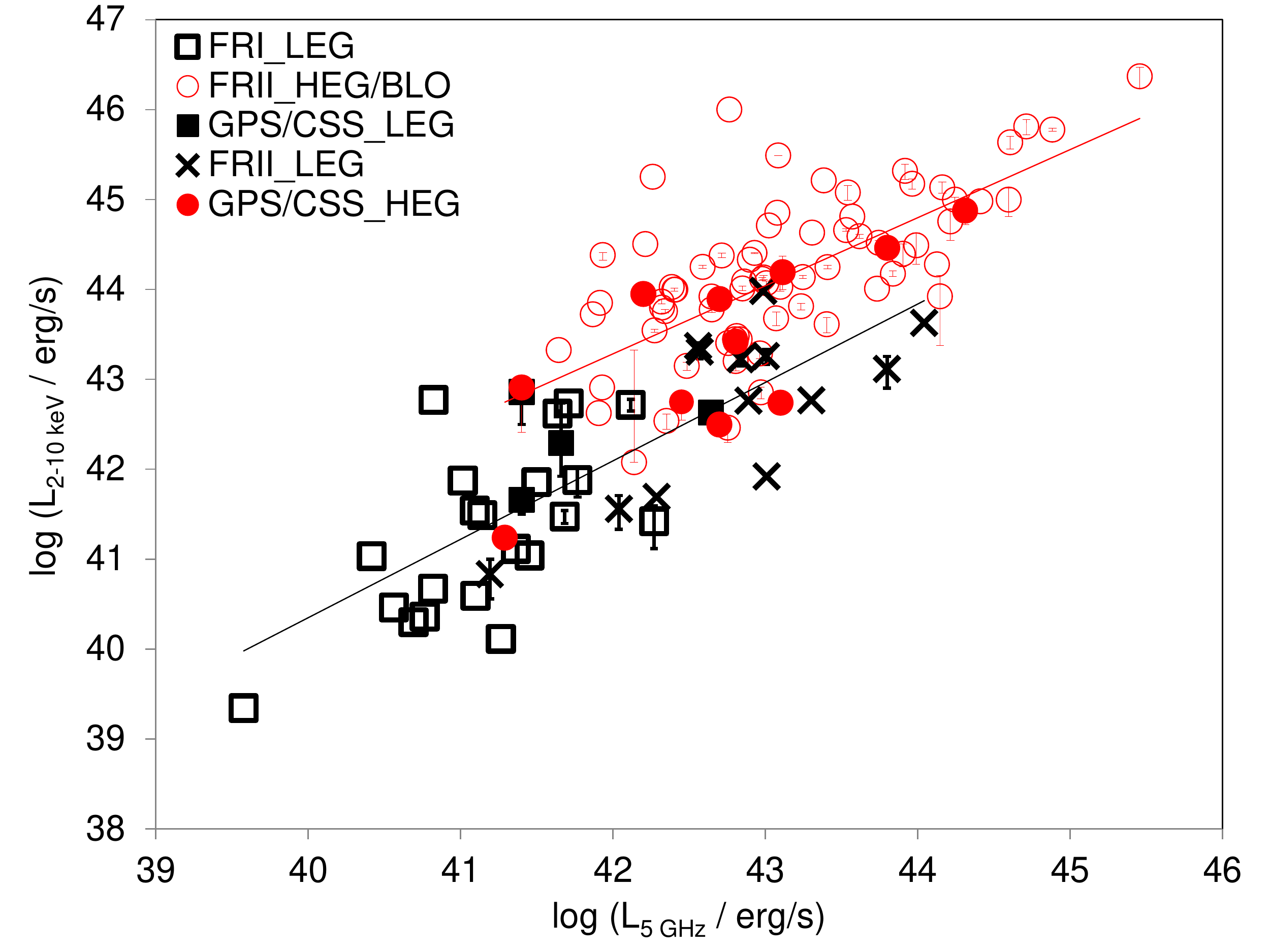}
\caption{2-10\,keV - 5\,GHz luminosity diagram for AGNs classified as HEGs and LEGs.
The FR\,II HEG/BLO sources and GPS/CSS HEG objects are indicated as open red
circles and red circles respectively. FR\,II LEGs are indicate with black   
crosses
and FR\,I LEGs and GPS/CSS LEGs with open and black squares respectively.}
\label{HEG_LEG_plot}
\end{figure}

The low power objects fit well to the already established X-ray -
radio luminosity correlation for AGNs and occupy the space among, weaker in
the X-rays, FR\,I objects. This trend is visible on both plots, X-ray vs.
356\,MHz and 5\,GHz
(Fig.\ref{5GHz_plot}), and is independent of radio frequency. However, the
356\,MHz radio luminosity versus X-ray luminosity plot shows larger scatter
among observable data than in the case of 5\,GHz luminosity. This is caused
by the fact that the X-ray emission is mostly associated with the compact
central regions of AGNs while the low frequency flux density is dominated by
the extended radio structures. 
As has been also shown by \citet{hard99} much of the dispersion in 5\,GHz
luminosity originate in beaming.    
Future X-ray observations of the whole sample of LLC
sources would give us a definitive information about their place on the
radio/X-ray luminosity plane.

\begin{table}
\caption{Correlation and Regression Analysis for Fig.~\ref{HEG_LEG_plot}}
\label{statistics}
\centering
\begin{tabular}{cccccc}
\hline
Sample   & N    &  
\multicolumn{1}{c}{r-Pearson}&
\multicolumn{2}{c}{\underline{~~~~~~~Linear Regression~~~~~~~}}\\
 & & 
\multicolumn{1}{c}{coefficient}& 
\multicolumn{1}{c}{Slope}& 
\multicolumn{1}{c}{Intercept}\\
\hline
HEG   & 82  &0.68 & $0.76\pm0.09$ &$11.49\pm3.90$ \\
LEG   & 38  &0.78 & $0.87\pm0.12$ &$5.45\pm4.81$  \\
\hline
\end{tabular}
\end{table}

We have then plotted all groups of AGNs on the 5\,GHz/X-ray luminosity plane
(Fig.~\ref{HEG_LEG_plot})
with a division for high excitation galaxies (HEG) and low excitation
galaxies (LEG). We took the optical identification from \citet{butti10} in the
case of FR\,I and FR\,II and indicated them as LEG and HEG/BLO. According to
\citet{butti10} the broad line objects (BLO) can be considered as members of
the HEG class. Identifications of GPS/CSS objects were taken from
\citep{Kunert10b} (see also Table~\ref{lumins} in this paper) and Table A1.   
We have only four LEGs among the GPS/CSS class and actually all of them have
been classified as CSS sources. HEGs are found among strong GPS and CSS
objects.  
The HEG/LEG plot confirms what we have previously found in
\citet{Kunert10b}. The HEG and LEG AGNs group in two different parts of the
plot. 

A Pearson correlation analysis applied to both sub-samples revealed a
significant X-ray/radio correlation (Table~\ref{statistics}).
In the radio versus X-ray luminosity plane (Fig.~\ref{5GHz_plot}), objects with a different
morphology are aligned
along the same correlation, with an increasing fraction of large-scale FR\,II
morphologies at higher
luminosities. Compact sources are well aligned along this correlation, with
weak CSS (strong CSS/GPS)
closer to the parameter space occupied by FR\,I (FR\,II). However, the
ionization mechanism seems to
discriminate more neatly radio sources in this plane. Low versus
high ionization sources occupy
distinct locus in this plane, notwithstanding their evolutionary stage. This
evidence agrees with a
scenario whereby the X-ray emission in large-scale HEG sources is dominated
by spectral components due
to (obscured) accretion as opposed to LEG objects where the X-ray emission
should be dominated by
non-thermal synchrotron jets \citep{hard09, antonucci, son12}. 
As seen on the radio/X-ray
luminosity diagram (Fig.~\ref{HEG_LEG_plot}), there are two branches, each
being
driven by different excitation mode and each containing compact sources.
The FR morphology, as well as the GPS and CSS
division, seems to be independent on the excitation modes \citep{butti10,  
Kunert10b, gendre13}.

\subsubsection{Radio/X-ray luminosity ratio}

\begin{figure*}
\centering
\includegraphics[width=8.5cm, height=6cm]{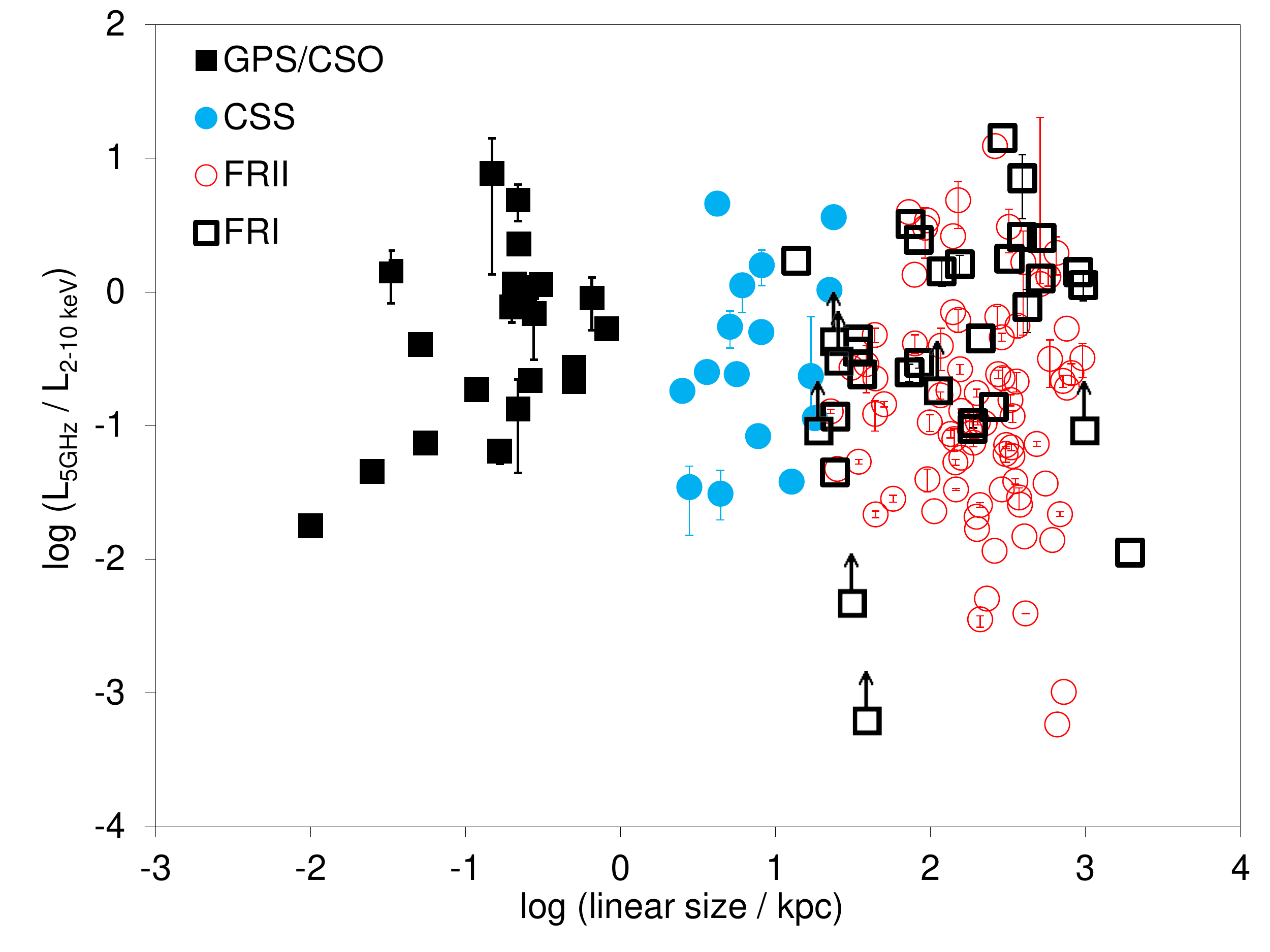}
\hfill
\includegraphics[width=8.5cm, height=6cm]{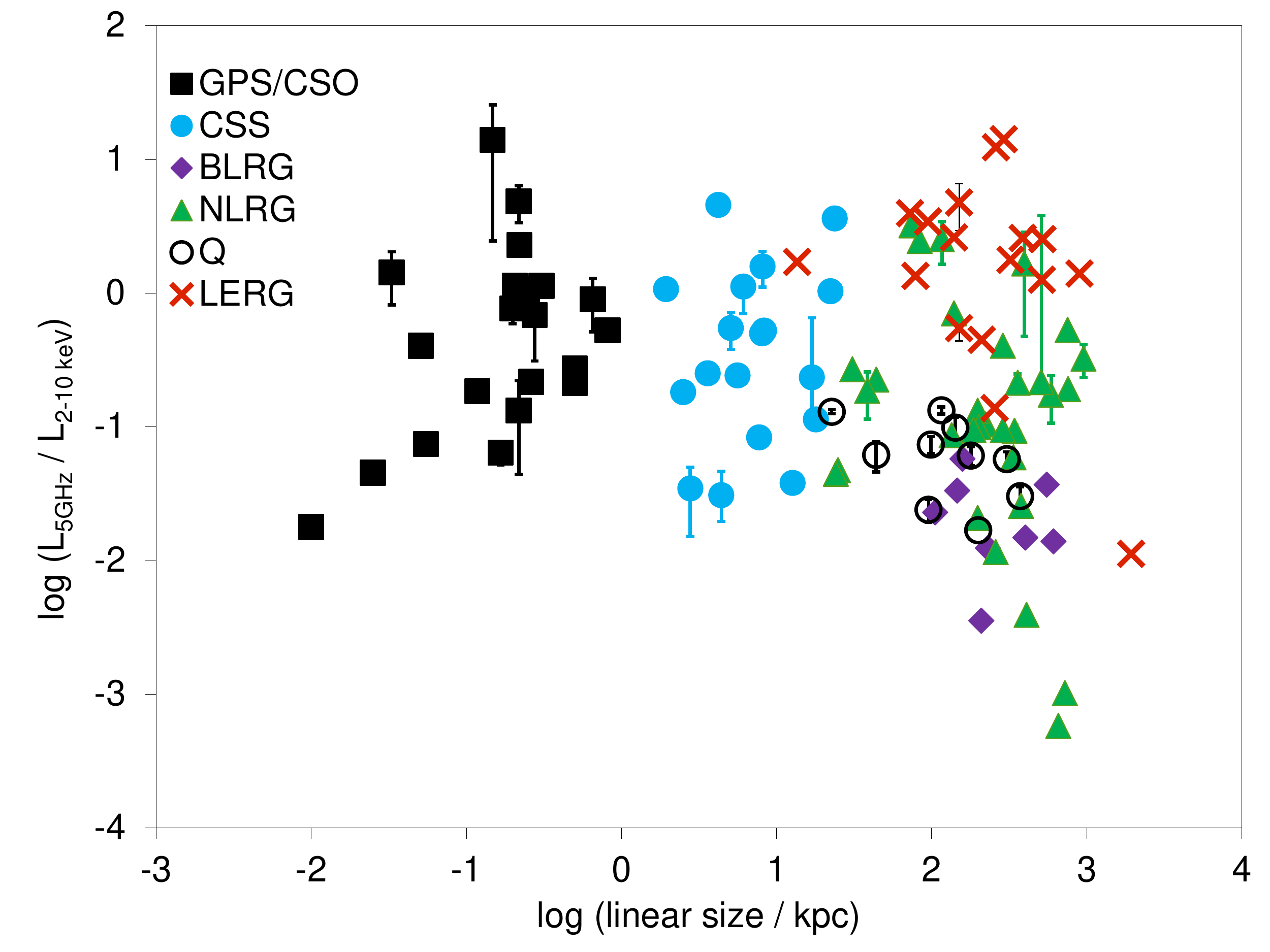}
\caption{The 5\,GHz luminosity/2-10\,keV luminosity ratio versus linear size
diagram for AGNs classified as follows. Left: GPS (black squares), CSS (blue  
circles), FR\,I (open black squares) and FR\,II (open red circles). Right: 
GPS (black squares), CSS (blue circles), LERG (red crosses), NLRG
(green triangles), Q (open black circles), BLRG (violet diamonds).}
\label{CSS_GPS_plot}
\end{figure*}

Another test for the AGN evolution models is a comparison of radio to the
X-ray luminosity ratio with the size of the sources, and indirectly, with
their age (Fig.~\ref{CSS_GPS_plot}). The long-term evolution of
extragalactic radio sources have been investigated by a number of authors in
different ways: i) as a variation of the radio power versus the total linear
size \citep{O'Dea97, Kunert10a, an12}, or ii) dynamic evolution of
FR\,II-like double radio sources characterized by the advance speed of the
hot spots, total extent of the source and depending on the density
distribution in the host galaxy along the path of the jets and lobes
\citep{Begelman89, Begelman96, Fanti95, Kaiser97b, Kino05, Kawakatu06, Kaiser07,
Kawakatu09a}. The radio luminosity of the sources evolves through
phases governed by the dominant energy-loss mechanism of the radiating,
relativistic electrons. From the onset of the radio activity, the
GPS/Compact Symmetric Object (CSO)
stage, the radio power of the sources increases with time and the source
size. The increase rate of radio power diminishes in the transition region 
(1-3 kpc from the center of
the host galaxy) where the balance between adiabatic losses and synchrotron
losses have been achieved. After this short period the radio power of CSS
sources starts to slowly decreases with the source size. The sharp decrease
in the radio power versus the total extent of the source occurs only in the large
FR\,I and FR\,II objects with the FR\,Is being below the luminosity
threshold, $L_{178 MHz} \sim$
$10^{25.5}$ W~H$z^{-1}$ $sr^{-1}$, on the radio power/linear size plane.
If the X-ray emission in radio-loud AGNs is due
to accretion only, the evolution of the radio and X-ray wavebands
could be totally decoupled and the radio/X-ray luminosity ratio should
reproduced the radio power evolution with the linear size of the source. The
Fig.~\ref{CSS_GPS_plot} shows that the above assumption is not
true not only in the case of large FR\,Is and FR\,IIs but probably also in
the case of young GPS and CCS sources. The less radio powerful FR\,Is
have the radio/X-ray luminosity ratio higher than many FR\,IIs what
may imply higher X-ray luminosity decrease with radio power in FR\,Is than
FR\,IIs. This can be explained by the idea that the X-ray emission in FR\,Is
originates from the base of a relativistic jet \citep{Evans06} and are thought
to be synchrotron emission \citep{Sambruna04, Worrall09}.
However, the X-ray emission of the FR\,IIs comes mostly from the obscured
X-ray component probably associated with the accretion and in less
part from the relativistic jet produced in an inverse Compton process
\citep{Balmaverde06, Belsole06, Evans06}.   
When incorporating a 
different division among large scale objects we notice that, on average, the
low excitation radio galaxies (LERG) and narrow line radio galaxies (NLRG)  
have higher radio to X-ray luminosity ratio than quasars (Q) and broad line
radio galaxies (BLRG). According to \cite{hard09} the common correlation of 
FR\,II NLRGs, LERGs and FR\,Is indicates that the X-ray and radio emission
comes from the same jet-related component.
The interpretation of the place of GPS and CSS sources on the radio/X-ray
luminosity ratio versus linear size plane is even more difficult because of
the large scatter of the observable values. As we have already mentioned the
beaming can disrupt the radio/X-ray correlations. However, at least
in the case of compact steep spectrum objects this effect should be small
\citep{wu13}.                  
We then suggest that what we observe in the group of young GPS and CSS
sources is a mix of two different types of X-ray/radio relation. We conclude
that at some radio power level the compact AGNs starts to resemble the FR\,Is,
where the X-ray emission is a synchrotron type associated with the jet.
It has been already proposed by the SED modelling of two strong CSS objects that
their X-ray emission can be a sum of X-ray emission from the accretion disk
and non-thermal X-ray emission from the parsec scale radio jet \citep{kunert2009, migliori12}. 

Recently, \citet{Stawarz08} and \citet{Ostorero10} have discussed an
alternative evolutionary model for GPS sources, which predicts the
dependency of the broadband Spectral
Energy Distribution on the source linear size.   
In their model high-energy  
emission is produced by upscattering of various photon fields by the lobes'
electrons of the $<$1\,kpc size radio source. This process can cause a decrease of the
X-ray to radio luminosity ratio by
1-2 orders of magnitude when the GPS source size increases.
However, with the data set gathered in this paper we cannot conclude
any correlation between the radio/X-ray luminosity ratio and linear size of
the sources. 

\subsubsection{$N_{H}$ - linear size relation}

\begin{figure}
\centering
\includegraphics[width=8.5cm, height=6cm]{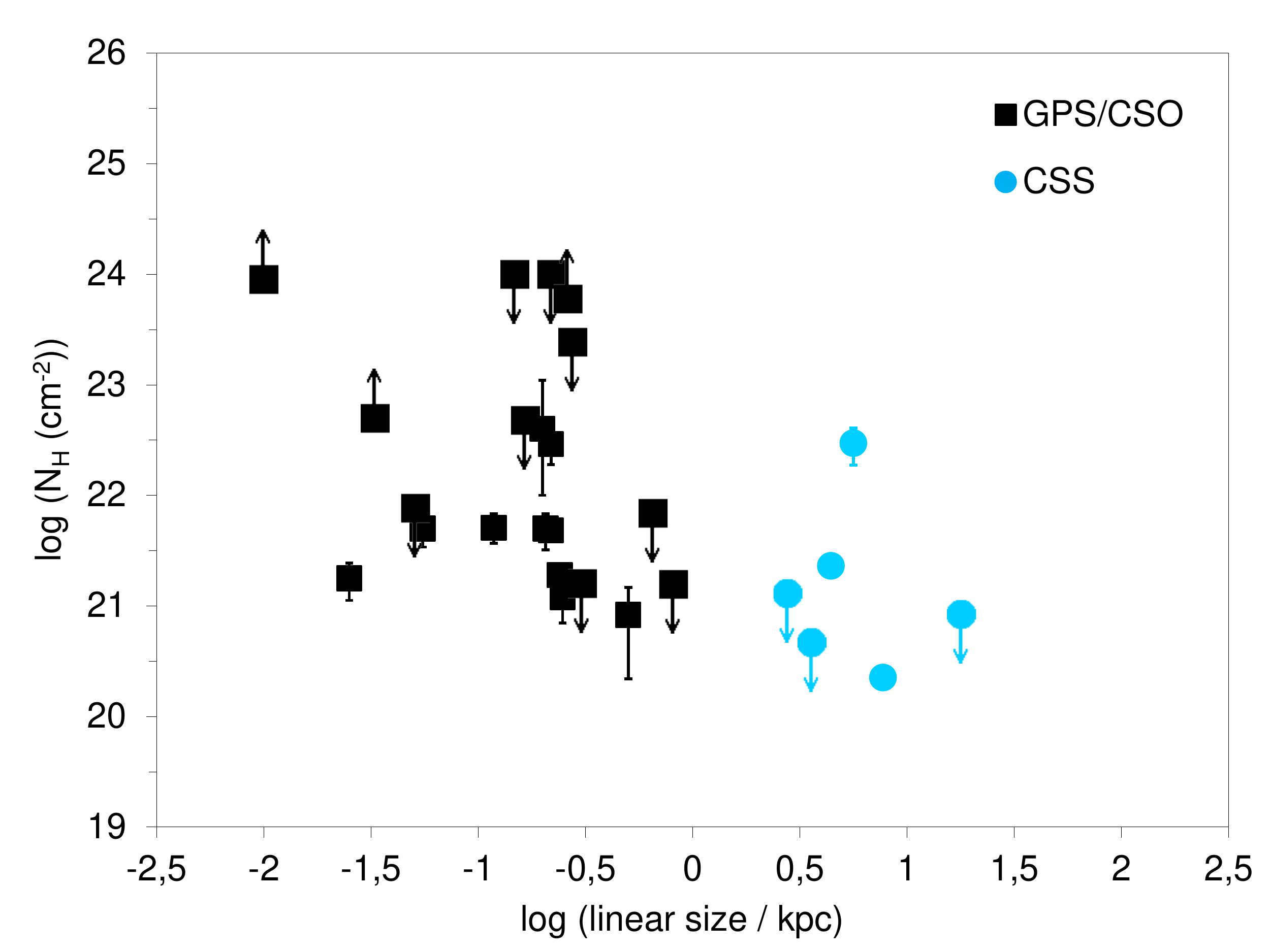}
\caption{X-ray column density versus linear size. GPS and CSS sources are
indicated as black squares and blue circles respectively.}
\label{N_H_plot}       
\end{figure}

Finally, we have drawn a relation between the measured column density and
the total extent of the radio source for GPS and CSS sources
(Fig.~\ref{N_H_plot}). It has been already reported \citep{Pihl03, Vermeulen03} 
that the small sources ($<$0.5\,kpc) tend
to have larger H\,I column density than larger sources ($>$0.5\,kpc) which
indicates that GPS/CSO objects evolve in a disk distribution of gas with a
power-law radial density dependence. The same explanation could lay behind
the (tentative)   
anti-correlation between X-ray column density and radio size in GPS galaxies
\citep{Tengstrand09}. \citet{Ostorero10} reported a positive correlation
between the radio and X-ray hydrogen column densities what can points toward
the cospatiality of the radio and X-ray emission regions. We extended the
discussion about the relation of the ${\rm N_{H}}$ versus the linear size
to the CSS sources. We have noticed that the ${\rm N_{H}}$ value of the CSS
sources is on average lower than that of GPS objects. However, the correlation
between the X-ray column density and radio size does not hold when including
larger CSS objects in the sample of small and young AGNs. 

\section{Summary}

In this paper we presented the X-ray {\it Chandra} observations of a pilot
sample of low luminosity compact steep spectrum (CSS) sources. Four of them
have been detected, the other three have upper limits estimations for the
X-ray flux. Only for two CSS objects we were able to estimate the X-ray column
density which is of the order of ${\rm 10^{21}\,cm^{-2}}$. We then expanded
the sample of compact AGNs with other GPS/CSS sources with X-ray detections 
found in the literature and used it, together with a sample of large FR\,I and FR\,II
sources, to determine the nature of the relation between morphology, X-ray
properties and excitation modes in radio-loud AGNs. We found the following
results:

$\bullet$
We have compared the X-ray luminosity of the radio sources from all above
mentioned groups with their radio properties. 
The large-scale FR\,II sources and strong GPS and CSS objects settle at
higher X-ray and radio luminosities.
While the low power CSSs occupy the space among, weaker in
the X-rays, FR\,I objects. This trend is visible independently of radio frequency.    

$\bullet$
The HEG and LEG sources occupy
distinct locus in the radio/X-ray luminosity plane, notwithstanding their
evolutionary stage. 
This is in agreement with the postulated different origin of the X-ray
emission in low and high ionization objects. Compact sources can be found in
both excitation modes driven branches.

$\bullet$
The less radio powerful FR\,Is
have higher radio/X-ray luminosity ratio than many FR\,IIs what may
imply higher X-ray luminosity decrease with radio power in FR\,Is than
FR\,IIs. This is in agreement with the previous findings saying that the X-ray emission in FR\,Is
originates from the base of a relativistic jet while the X-ray emission of
FR\,IIs has accretion origin. The same can be true in the case of smaller
radio AGNs, namely the GPS and CSS sources. The result of this study hints toward
the fact that below some radio power level the compact GPS and CSS sources
start to resemble the FR\,Is, or to be more specific, the LERG and NLRG objects.

$\bullet$
The X-ray hydrogen column density of the CSS
sources is on average lower than that of GPS objects. But the
correlation
between the X-ray column density and radio size does not hold for the whole
sample of GPS and CSS objects.

\section*{Acknowledgements}
AL wishes to thank CfA for their hospitality and support during the visit.

This research has made use of data obtained by the Chandra X-ray
Observatory, and {\it Chandra} X-ray Center (CXC) in the application
packages CIAO, ChIPS, and Sherpa.  This research is funded in part by
(NASA) contract NAS8-03060. Partial support for this work was provided
by the {\it Chandra} grants GO1-12124X and GO1-12145X.

This research has made use of NASA's Astrophysics Data System Bibliographic 
Services and of the NASA/IPAC Extragalactic Database (NED) which is operated 
by the Jet Propulsion Laboratory, California Institute of Technology, 
under contract with the National Aeronautics and Space Administration. 

This publication makes use of data products from the SDSS.
    Funding for the SDSS and SDSS-II has been provided by the Alfred P.
Sloan Foundation, the Participating Institutions, the National Science 
Foundation, the U.S. Department of Energy, the National Aeronautics and
Space Administration, the Japanese Monbukagakusho, the Max Planck Society,
and the Higher Education Funding Council for England. The SDSS Web Site is
http://www.sdss.org/.

    The SDSS is managed by the Astrophysical Research Consortium for the
Participating Institutions. The Participating Institutions are the American
Museum of Natural History, Astrophysical Institute Potsdam, University of  
Basel, University of Cambridge, Case Western Reserve University, University
of Chicago, Drexel University, Fermilab, the Institute for Advanced Study, 
the Japan Participation Group, Johns Hopkins University, the Joint Institute
for Nuclear Astrophysics, the Kavli Institute for Particle Astrophysics and 
Cosmology, the Korean Scientist Group, the Chinese Academy of Sciences
(LAMOST), Los Alamos National Laboratory, the Max-Planck-Institute for
Astronomy (MPIA), the Max-Planck-Institute for Astrophysics (MPA), New
Mexico State University, Ohio State University, University of Pittsburgh,
University of Portsmouth, Princeton University, the United States Naval  
Observatory, and the University of Washington.

\appendix

\section[]{Spectroscopic classification of the other CSS/GPS sources with
X-ray data}

The spectroscopic data are available for 6 CSS sources from the
samples of CSS/GPS sources of \citet{Tengstrand09} and
\citet{Siemiginowska08} (Table A1). The HEG/LEG 
classification was based on the line ratios observed in the SDSS spectra according to the
description by \citet{Kunert10b}.

\begin{table*}
\caption[]{Emission lines measurements and spectroscopic classification of
compact sources. The columns show the following : (1) source name, (2)-(3) coordinates, (4) redshift, (5)-(7) flux in $\rm
10^{-17}~erg~s^{-1}~cm^{-2}$, O\,II - total flux of the O\,II doublet, (8)-(10) log of
luminosities; luminosities are in erg s\mone, (11)
spectroscopic classification. Wavelengths are in Angstroms. 'a' means the
classification is based on the \oo \, ratio only, 'b' see also
\citet{Labiano05}.}

\begin{tabular}{@{}l c c c c c c c c c l @{}}
\hline
Source name& RA(J2000) & Dec (J2000) & {\it z} &  
\multicolumn{1}{c}{[O II]} &
\multicolumn{1}{c}{[O III]} &
\multicolumn{1}{c}{H$_\beta$} &
log L$_{2-10 \rm{keV}}$ & log L$_{5 \rm{GHz}}$ & log L$_{\rm[{O}{III}]}$ &
\multicolumn{1}{c}{Spectral}\\
   & hh mm ss  & dd mm ss    &        &
\multicolumn{1}{c}{$\lambda 3727$}&
\multicolumn{1}{c}{$\lambda 5007$}& 
\multicolumn{1}{c}{$\lambda 4861$}&
 & & &
\multicolumn{1}{c}{type}\\
 & & & & 
\multicolumn{1}{c}{$\lambda 3729$}&
 & & & & & \\
\hline
B2 0738+31 & 07:41:10.7 & 31:12:00 & 0.63 & 445.7 & 1490.6 & 3675.6 & 44.9 &
44.3&43.4  & ${\rm HEG^{a}}$\\
Q1250+568  & 12:52:26.3 & 56:34:20 & 0.32 & 1066.0 & 3564.8 & 1653.3 & 44.2& 
43.2&43.1  & ${\rm HEG^{b}}$\\
1345+125   & 13:47:33.3 & 12:17:24 & 0.12 & 1038.9 & 4682.3 & 623.9  & 43.8&
42.7&42.2  & HEG\\
OQ+208     & 14:07:00.4 & 28:27:15 & 0.07 & 37.8   & 1068.0 & 18749.0 &
$>$43.9 & 42.2& 41.2& ${\rm HEG^{a}}$\\
4C+00.02   & 00:22:25.4 & 00:14:56 & 0.30 & 29.2   & 54.2   & 2.4 & 42.7&
43.1&  41.2& HEG\\
PKS1607+26 & 16:09:13.3 & 26:41:29 & 0.47 & 63.6   & 200.0  & 44.6& 44.5&
43.8& 42.2& HEG\\
\hline
\end{tabular}
\end{table*}

\end{document}